\documentclass[letterpaper]{article} % DO NOT CHANGE THIS
\usepackage{aaai2026}  % DO NOT CHANGE THIS
\usepackage{times}  % DO NOT CHANGE THIS
\usepackage{helvet}  % DO NOT CHANGE THIS
\usepackage{courier}  % DO NOT CHANGE THIS
\usepackage[hyphens]{url}  % DO NOT CHANGE THIS
\usepackage{graphicx} % DO NOT CHANGE THIS
\usepackage{natbib}  % DO NOT CHANGE THIS AND DO NOT ADD ANY OPTIONS TO IT
\usepackage{caption} % DO NOT CHANGE THIS AND DO NOT ADD ANY OPTIONS TO IT
\usepackage{algorithm}
\usepackage{algorithmic}

\usepackage{newfloat}
\usepackage{listings}
\DeclareCaptionStyle{ruled}{labelfont=normalfont,labelsep=colon,strut=off} % DO NOT CHANGE THIS
\floatstyle{ruled}
\newfloat{listing}{tb}{lst}{}
\floatname{listing}{Listing}

\usepackage{amssymb}
\usepackage{multirow}
\usepackage{booktabs}
\usepackage{pifont}

\newcommand{\ie}{\textit{i}.\textit{e}.}
\newcommand{\eg}{\textit{e}.\textit{g}.}
\newcommand{\score}[2]{{#1}\scriptsize{$\pm$#2}}
\newcommand{\cmark}{\ding{51}}%
\usepackage{newfloat}
\usepackage{listings}
\DeclareCaptionStyle{ruled}{labelfont=normalfont,labelsep=colon,strut=off} % DO NOT CHANGE THIS
\floatstyle{ruled}
\newfloat{listing}{tb}{lst}{}
\floatname{listing}{Listing}
\title{Towards Effective and Efficient Context-aware Nucleus Detection in Histopathology Whole Slide Images}
\author{
    Zhongyi Shui\textsuperscript{\rm 1,2}\equalcontrib,
    Honglin Li\textsuperscript{\rm 1,2}\equalcontrib,
    Yunlong Zhang\textsuperscript{\rm 1,2},
    Yuxuan Sun\textsuperscript{\rm 1,2},
    Yiwen Ye\textsuperscript{\rm 3},
    Pingyi Chen\textsuperscript{\rm 1,2},
    Ruizhe Guo\textsuperscript{\rm 1,2},
    Lei Cui\textsuperscript{\rm 4},
    Chenglu Zhu\textsuperscript{\rm 2$\dagger$},
    Lin Yang\textsuperscript{\rm 2,5,6}\thanks{Corresponding author.}
}
\affiliations{
    \textsuperscript{\rm 1} College of Computer Science and Technology, Zhejiang University \\
    \textsuperscript{\rm 2} School of Engineering, Westlake University\\
    \textsuperscript{\rm 3} College of Computer Science and Technology, Northwestern Polytechnical University\\
    \textsuperscript{\rm 4} School of Information Science and Technology, Northwest University\\
    \textsuperscript{\rm 5} The Institute of Advanced Technology, Westlake Institute for Advanced Study\\
    \textsuperscript{\rm 6} Center for Interdisciplinary Research and Innovation, MuyuanLaboratory
}

\usepackage{bibentry}
\begin{document}
	
\maketitle

\begin{abstract}
	Nucleus detection in histopathology whole slide images (WSIs) is crucial for a broad spectrum of clinical applications. The gigapixel size of WSIs necessitates the use of sliding window methodology for nucleus detection. However, mainstream methods process each sliding window independently, which overlooks broader contextual information and easily leads to inaccurate predictions. To address this limitation, recent studies additionally crop a large Filed-of-View (LFoV) patch centered on each sliding window to extract contextual features. However, such methods substantially increase whole-slide inference latency. In this work, we propose an effective and efficient context-aware nucleus detection approach. Specifically, instead of using LFoV patches, we aggregate contextual clues from off-the-shelf features of historically visited sliding windows, which greatly enhances the inference efficiency. Moreover, compared to LFoV patches used in previous works, the sliding window patches have higher magnification and provide finer-grained tissue details, thereby enhancing the classification accuracy.
	To develop the proposed context-aware model, we utilize annotated patches along with their surrounding unlabeled patches for training. Beyond exploiting high-level tissue context from these surrounding regions, we design a post-training strategy that leverages abundant unlabeled nucleus samples within them to enhance the model's context adaptability. Extensive experimental results on three challenging benchmarks demonstrate the superiority of our method.
\end{abstract}

\begin{links}
\link{Code}{https://github.com/windygoo/PathContext}
\end{links}

\section{Introduction}

\begin{figure}[t!]
	\centering
	\includegraphics[width=\linewidth]{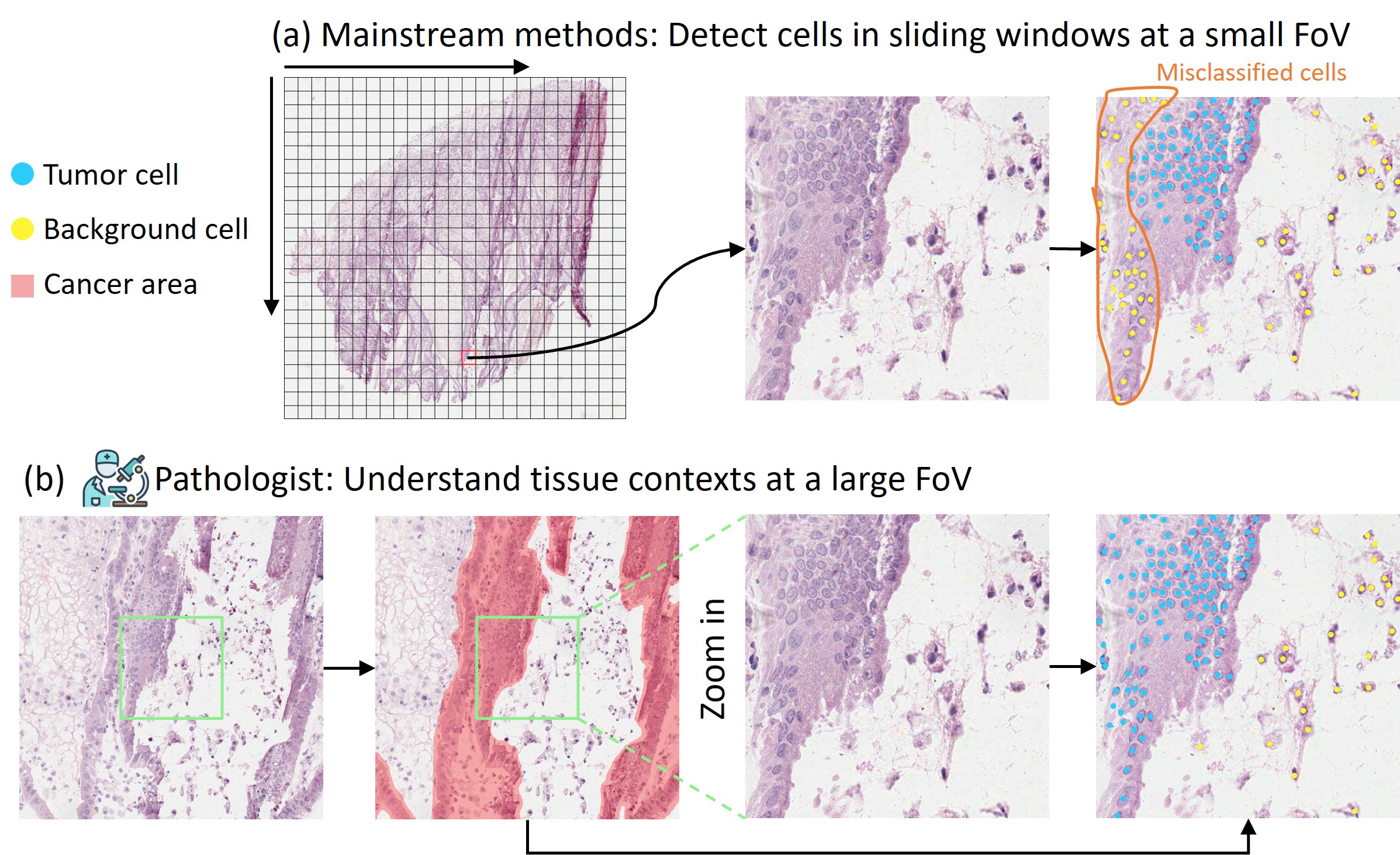}
	\caption{Nucleus detection on gigapixel WSIs necessitates a sliding window strategy. (a) Mainstream methods detect nuclei in each window patch without understanding broader tissue structure, which easily leads to inaccurate predictions. (b) Pathologists first zoom out to examine tissue context at large FoVs and then zoom in to observe detailed nuclear morphology for accurate nucleus classification \cite{ryu2023ocelot}.}\label{fig:back}
\end{figure}

Nucleus detection in histopathology whole slide images (WSIs) is a fundamental task in computational pathology. It allows quantitative analysis of WSIs, which can lead to better cancer diagnosis, grading, prognosis, and treatment planning while maintaining medical interpretability \cite{diao2021human,ryu2023ocelot,yang2024histopathology,ignatov2024histopathological,xu2025co}. Therefore, the development of precise automatic nucleus detection algorithms has become a critical research focus in recent years.
Current nucleus detection pipeline involves training a nucleus detector using expert-annotated histopathology patches and then deploys it to detect nuclei in gigapixel WSIs through a sliding window technique \cite{huang2020bcdata,shui2022end,zhang2022weakly,huang2023affine}, as illustrated in Fig.~\ref{fig:back} (a). However, for accurate nucleus localization, the annotated and sliding window patches are cropped at high magnification but small Field-of-View (FoV) \cite{ryu2023ocelot}. As a result, the nucleus detector can only see a limited context without understanding broader tissue information, which can easily lead to inaccurate predictions. In clinical practice, pathologists first examine tissue context at large FoVs and then zoom in to observe detailed nuclear morphology for accurate assessments, as depicted in Fig.~\ref{fig:back} (b).

\begin{figure}[t!]
	\centering
	\includegraphics[width=\linewidth]{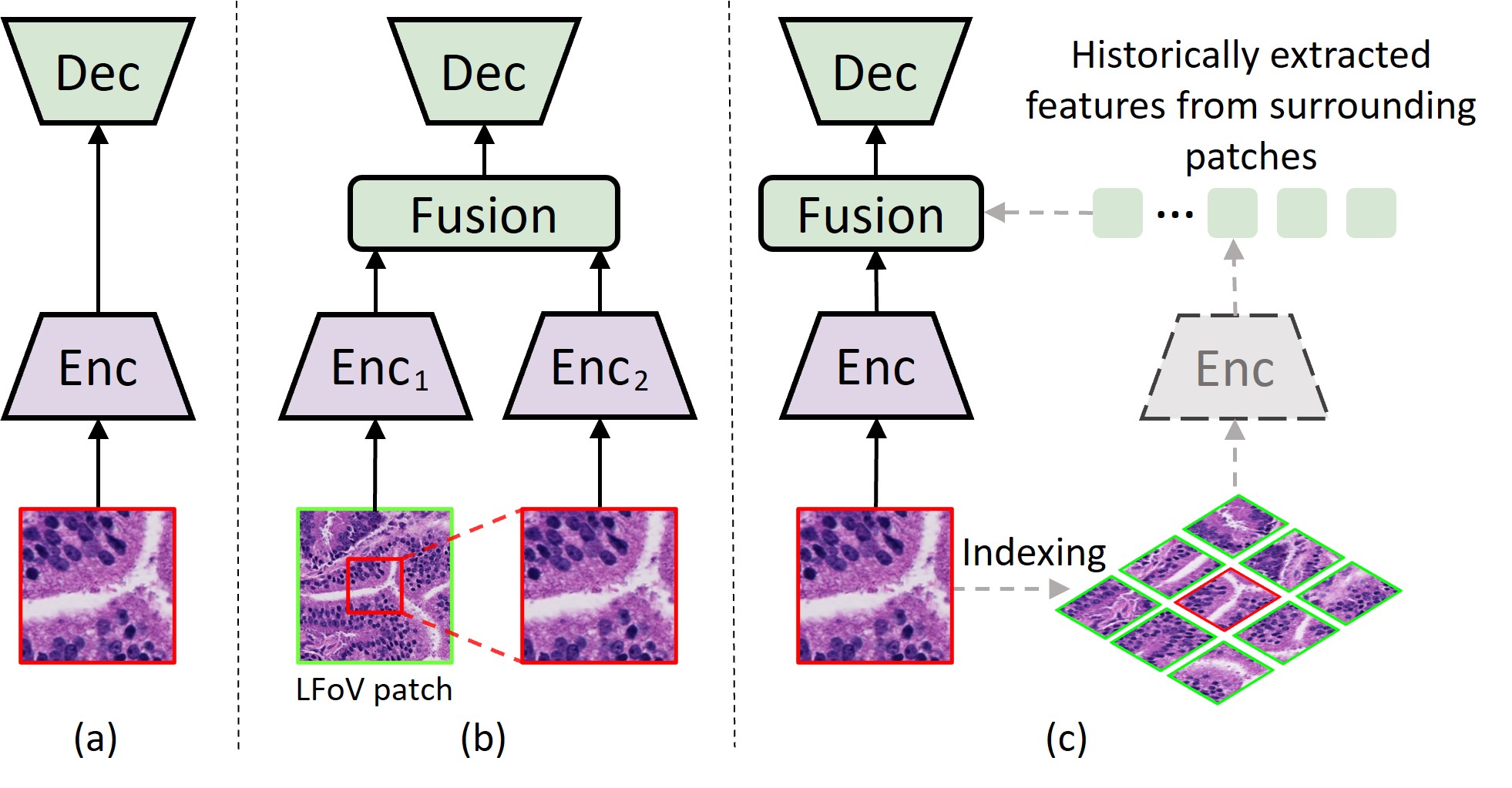}
	\caption{(a) Mainstream nucleus detection methods operate on patches at a single FoV without considering tissue context. (b) Previous context-aware nucleus detection approaches leverage large FoV (LFoV) patches to extract contextual information, which substantially increases the whole-slide inference time due to additional I/O-intensive data preparation. (c) The proposed context-aware method aggregates contextual information from off-the-shelf features of historically visited window patches, which greatly improves the whole-slide inference efficiency.}\label{fig:banner}
\end{figure}

Inspired by this clinical workflow, recent works \cite{bai2020multi,bai2022context,shui2024dpa,ryu2023ocelot,millward2023dense,torbati2025multi} utilize a low-magnification large FoV (LFoV) patch, which has the same size as sliding window patches but encompass broader WSI regions, as a supplementary input of nucleus detectors to enable context-aware nucleus detection, as shown in Fig.~\ref{fig:banner} (b). Despite the improved outcomes, we argue that this line of approaches exhibits two critical limitations: (1) The requirement for additional processing the LFoV patches significantly increases the whole-slide inference time. (2) The LFoV patches at low magnification inherently lack fine-grained tissue details, thereby diminishing the potential performance gains.

In this work, we propose to aggregate contextual information from patches that surround with the  region-of-interest (ROI) patch, as illustrated by Fig.~\ref{fig:banner} (c). Notably, in the training stage, the ROI patch comes from the annotated set while during inference on WSIs, the ROI patch and its surrounding patches are both part of sliding windows. Therefore, this design eliminates the I/O-intensive and time-consuming step of additionally preparing LFoV patches in previous studies. Moreover, since the ROI patch and its surrounding patches have the same magnification (\ie, data distribution), we employ a shared image encoder to process both and utilize features extracted from surrounding patches as contextual clues. With this design, we can directly re-use the off-the-shelf features extracted from historically visited surrounding windows to perform context-aware nucleus detection, which further improves the whole-slide inference speed.

Additionally, we observe that current context-aware methods exclusively exploit high-level contextual features in LFoV patches while neglecting the massive unlabeled nuclei within these patches. To harness this untapped resource, we introduce a post-training stage in this work. Specifically, we employ the pre-trained detector to detect these unlabeled nuclei and generate pseudo labels for them using a novel cross-labeling strategy. Then, we use these pseudo-labeled samples to fine-tune the detector, empowering it to classify nuclei at different spatial locations (\ie, various context conditions) and thus enhancing the model's context adaptability. Besides, we discover for the first time that incorporating high-level contextual features inherently diminishes the model's perception of low-level nuclear morphological details, which potentially comprises the model's accuracy. To address this limitation, we introduce an lightweight auxiliary branch to compensate for these morphological features. Our main contributions can be summarized as follows:
\begin{enumerate}
	\item We propose a novel context aggregation approach that exploits features from surrounding sliding windows for effective and efficient nucleus detection in WSIs.
	\item We propose a cross-labeling strategy to effectively utilize unlabeled nuclei in surrounding patches to improve the model's context adaptability.
	\item Extensive experiments on three challenging benchmarks demonstrate the advantages of our method over the state-of-the-art counterparts on both nucleus detection and instance segmentation tasks.
\end{enumerate}

\begin{figure*}[t!]
	\centering
	\includegraphics[width=\linewidth]{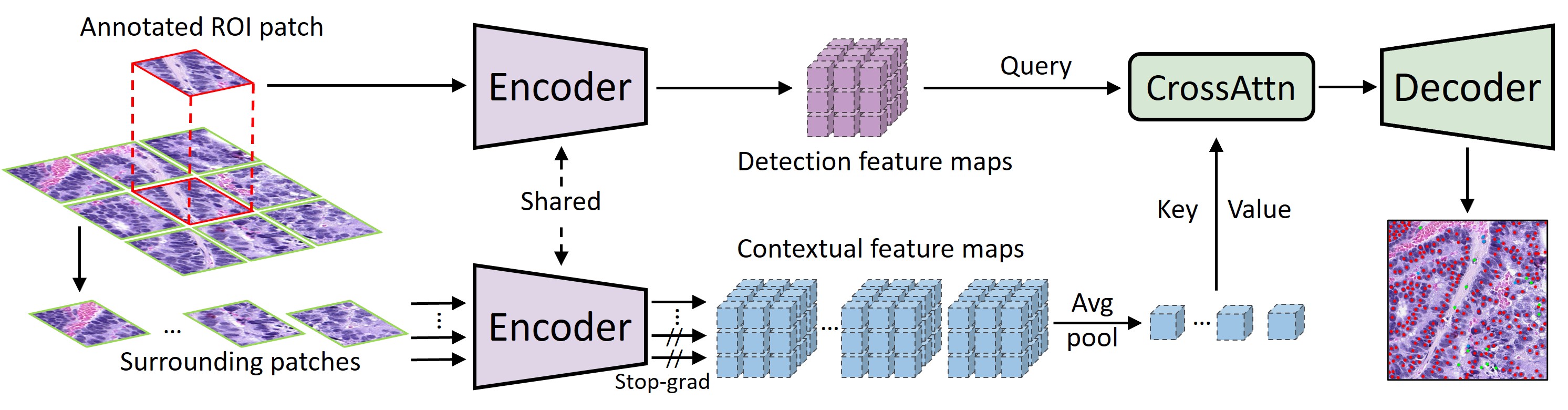}
	\caption{The training pipeline of our proposed context-aware nucleus detection method. We use a shared visual encoder to encode the annotated patch and its surrounding patches. To avoid GPU memory overflow from simultaneously processing numerous surrounding patches with gradient computation, we randomly select a small subset to participate in back-propagation while the rest undergo forward pass only. The contextual features are first downsampled through grid average pooling and then incorporated into the detection branch via cross-attention.
		%	Notably, we apply grid average pooling to downsample the contextual feature maps, thereby eliminating their spatial redundancy \cite{li2025pathvq} and enabling the model to capture informative contextual features in cross-attention.
	}\label{fig:framework}
\end{figure*}

\section{Related Works}

\subsection{Nucleus Detection}\label{sec:nuc_det}
Current methods for nucleus detection can be divided into two categories: density map-based and end-to-end. 

Density map-based methods \cite{graham2019hover,abousamra2021multi,zhang2022weakly,pan2023smile,lou2024cell,horst2024cellvit} first regress nucleus probability maps, and then apply post-processing including thresholding, local maxima detection, and non-maximum suppression to identify nuclei centroids.  In contrast, end-to-end methods can directly predict nuclear positions without hand-crafted post-processing procedures. These approaches can be further categorized into two distinct paradigms: anchor-based and anchor-free. Anchor-based methods, primarily built upon P2PNet \cite{song2021rethinking}, localize nuclei by predicting relative offsets from pre-defined anchor points across input images \cite{song2021rethinking,shui2022end,shui2024dpa}, while anchor-free approaches \cite{huang2023affine,huang2023prompt,pina2024cell} directly predict absolute nuclear positions through learnable queries following the DETR architecture \cite{carion2020end}. Due to the substantial variability in nuclei size, shape and spatial distribution, it is impossible to design a universal post-processing approach that generalizes well across all scenarios. Consequently, density map-based methods generally exhibit inferior detection performance and efficiency compared to end-to-end approaches.
% Furthermore, among end-to-end algorithms, anchor-based methods typically demonstrate superior performance compared to anchor-free counterparts, primarily attributed to their inherent inductive biases that provide favorable spatial priors for nucleus localization \cite{shui2024unleashing}.

\subsection{Context-aware Pathology Image Analysis}
Due to the gigapixel scale of WSIs, dense prediction tasks such as tissue segmentation and nucleus detection on them must be conducted in a sliding window manner. However, this strategy can lead to the loss of broader spatial context when window patches are processed independently. To address this issue, several recent works have explored context-aware approaches for tissue segmentation \cite{kamnitsas2017efficient,tokunaga2019adaptive,schmitz2021multi,van2021hooknet} and nucleus detection \cite{bai2020multi,bai2022context,ryu2023ocelot,shui2024dpa}. 

Current context-aware nucleus detection approaches can be divided into two categories: explicit and implicit. Explicit approaches \cite{ryu2023ocelot,schoenpflug2023softctm,torbati2025multi} rely on additional tissue region annotations. Specifically, they utilize an auxiliary model for tissue segmentation at LFoV, and the predicted tissue mask is fed into the detection model to enhance nucleus classification accuracy. In contrast, the latter approaches \cite{bai2020multi,bai2022context,shui2024dpa} eliminate the need for tissue mask labels and learn context features from LFoV patches implicitly, as illustrated in Fig.~\ref{fig:banner} (b). Despite notable performance improvements, all these methods introduce LFoV patches as additional input, which significantly increases the inference overhead. Essentially, the method proposed in this work belongs to the implicit family. However, different from previous works that utilize LFoV patches, we aggregate contextual information from off-the-shelf features of historically visited sliding window patches, which greatly improves the whole-slide inference efficiency.

\section{Method}
\subsection{Background}
Mainstream nucleus detectors takes a single patch as input \cite{graham2019hover,shui2024dpa,horst2024cellvit}. Such models are trained on an annotated patch set $\mathcal{D}=\{(x_i,p_i,y_i)\}_{i=1}^{N}$. For each patch $x_i \in \mathbb{R}^{H\times W\times 3}$, the annotation $p_i$ and $y_i$ represent centroids and categories of all nuclei in it, respectively. To develop context-aware nucleus detector, each annotated patch $x_i$ is complemented by its surrounding unlabeled patches $\{x_{i,j,k} \in \mathbb{R}^{H\times W\times 3}\mid j, k \in \{-\delta, \ldots, \delta\}\}$, where $\delta$ denotes the size of context area considered by the detector. For instance, when $\delta=1$, the model learns to detect nuclei in an annotated patch while  considering tissue context from its corresponding $3\times 3$ neighborhood. Following \cite{shui2024dpa}, we adopt P2PNet, an end-to-end nucleus detector, as the base model. Fig.~\ref{fig:framework} depicts the training pipeline of our proposed method.

%For instance, when $\delta$ equals 1, the model learns to detect nuclei in an annotated patch while aggregating contextual information from its eight adjacent patches, as illustrated in Fig.~\ref{fig:framework}.

\subsection{Extraction of Context Features}
Current context-aware nucleus detection works \cite{bai2020multi,bai2022context,shui2024dpa,ryu2023ocelot} all leverage LFoV patches to extract contextual information that improves nucleus detection accuracy in ROI patches at small FoV. As these two types of patches have different magnifications and thereby lie in distinct data distributions \cite{chen2022self}, these approaches employ separate image encoders to process them. Differently, since we extract contextual features from patches that surround the ROI patch and they have the same magnification, a shared image encoder is employed in this work.

During training, we encode an annotated patch $x_i$ into $\mathcal{F}_i \in \mathbb{R}^{h\times w\times d}$ and its surrounding patches into context feature maps $\{\mathcal{F}_{i,j,k} \in \mathbb{R}^{h\times w\times d}\mid j, k \in \{-\delta, \ldots, \delta\}\}$. Unlike previous methods that involve only two patches in each iteration, our approach requires encoding $(2\delta+1)^2$ patches concurrently. When $\delta=1$, this amounts to 9 patches, and enabling gradient computation for all of them leads to prohibitive memory requirements. To address this challenge, we propose a selective gradient computation strategy. Specifically, in each iteration, we randomly select $k$ surrounding patches for back-propagation while the rest $(2\delta+1)^2-k$ patches undergo feature extraction in the gradient-free manner. This approach preserves the model's capability to capture informative contextual features while significantly reducing memory consumption. 

% and thus helping the model capture informative contextual information
To eliminate spatial redundancy \cite{li2025pathvq} in each context feature map, we downsample $\mathcal{F}_{i,j,k}$ by partitioning it into a uniform $s\times s$ grid and apply average pooling within each grid cell. This reduces the resolution of $\mathcal{F}_{i,j,k}$ from $hw$ to $s^2$, where $s \ll \min(h,w)$ is set empirically in this work. Finally, we concatenate all compressed context feature maps as $\mathcal{F}_{i}^{ctx} \in \mathbb{R}^{(2\delta+1)^2\times s\times s\times d}$.

\subsection{Injection of Context Features}
To enable context-aware nucleus detection, we inject  the context feature maps $\mathcal{F}_{i}^{ctx}$ into the hidden embedding $\mathcal{F}_i$ extracted from the annotated patch via cross-attention:
\begin{equation}\label{eq:cross}
	\mathcal{F}_{i}^\prime = {\rm CrossAttn}\left(\rm{Q}=\mathcal{F}_{i}, \rm{K}=\mathcal{F}_{i}^{ctx}, \rm{V}=\mathcal{F}_{i}^{ctx}\right)
\end{equation}where $\mathcal{F}_{i}$ serves as query and $\mathcal{F}_{i}^{ctx}$ serves as both key and value. Finally, the context-enriched $\mathcal{F}_{i}^\prime$ is fed into the decoder to predict nucleus centroids and categories in the annotated patch. It is worth noting that we observe no performance gains when adding positional embeddings in Eq.~\ref{eq:cross}. We hypothesize that this is because histopathology slides are inherently continuous, with adjacent patches displaying coherent visual content along their boundaries. This continuity implicitly encodes relative positional relationship.

\subsection{Enhancing Context Adaptability with Unlabeled Nuclei}
We observe that the LFoV patches or surrounding patches provide not only high-level tissue context but also abundant unlabeled nucleus samples. To leverage this resource, we introduce a post-training stage that enables the model to classify nuclei at different spatial locations (\ie, various context conditions) to enhance its context adaptability.

In general, the proposed context-aware nucleus detection method can be decomposed into two steps: (1) generating context-enriched embedding $e\in\mathbb{R}^d$ for each point proposal \cite{song2021rethinking} and (2) performing classification via a classifier head $\phi: e \to Y\in\mathbb{R}^{C+1}$, where $C$ represents the number of nucleus categories and the extra class is background. For each annotated patch $x_i$, we utilize the nucleus detector pre-trained in the above sections to identify nuclei in its surrounding patches $\{x_{i,j,k} \mid j, k \in \{-\delta, \ldots, \delta\}\}$ while concurrently generating pseudo class labels for them. This yields additional training data $\{(x_{i,j,k},\hat{p}_{i,j,k},\hat{y}_{i,j,k}) \mid j, k \in \{-\delta, \ldots, \delta\}\}$ along with context-enriched embeddings $e$ for each identified nucleus. Then, we train a MLP head $\phi^\prime: e \to Y^\prime \in\mathbb{R}^{C}$, where the embeddings $e$ corresponding to nuclei in surrounding patches are supervised by pseudo labels $\hat{y}_{i,j,k}$ while those from the annotated patch use ground-truth class labels $y_i$.

However, we observe that using pseudo labels predicted by the nucleus detector itself leads to marginal improvements, even when we apply a high confidence threshold (\eg, 0.9) to filter out unreliable pseudo labels. This can be attributed to the confirmation bias inherent in self-training approaches \cite{arazo2020pseudo}, which leads to error accumulation when the model's own predictions are used for self-supervision. To mitigate this problem, we propose a cross-labeling technique. Specifically, we convert point annotations to pseudo mask labels following \cite{lin2024bonus} and train a lightweight auxiliary model for multi-class nucleus segmentation. Afterwards, we feed each surrounding patch $x_{i,j,k}$ into it and extract pseudo labels $\hat{y}_{i,j,k}$ for all pre-detected nuclei according to their coordinates $\hat{p}_{i,j,k}$ on the predicted class maps. We find although the auxiliary model shows only comparable classification accuracy to our detector, the substantial difference on model architecture (density map-based vs. end-to-end) and training regime leads to distinct classification patterns and greatly alleviates the error accumulation problem.

\begin{figure}[t!]
	\centering
	\includegraphics[width=\linewidth]{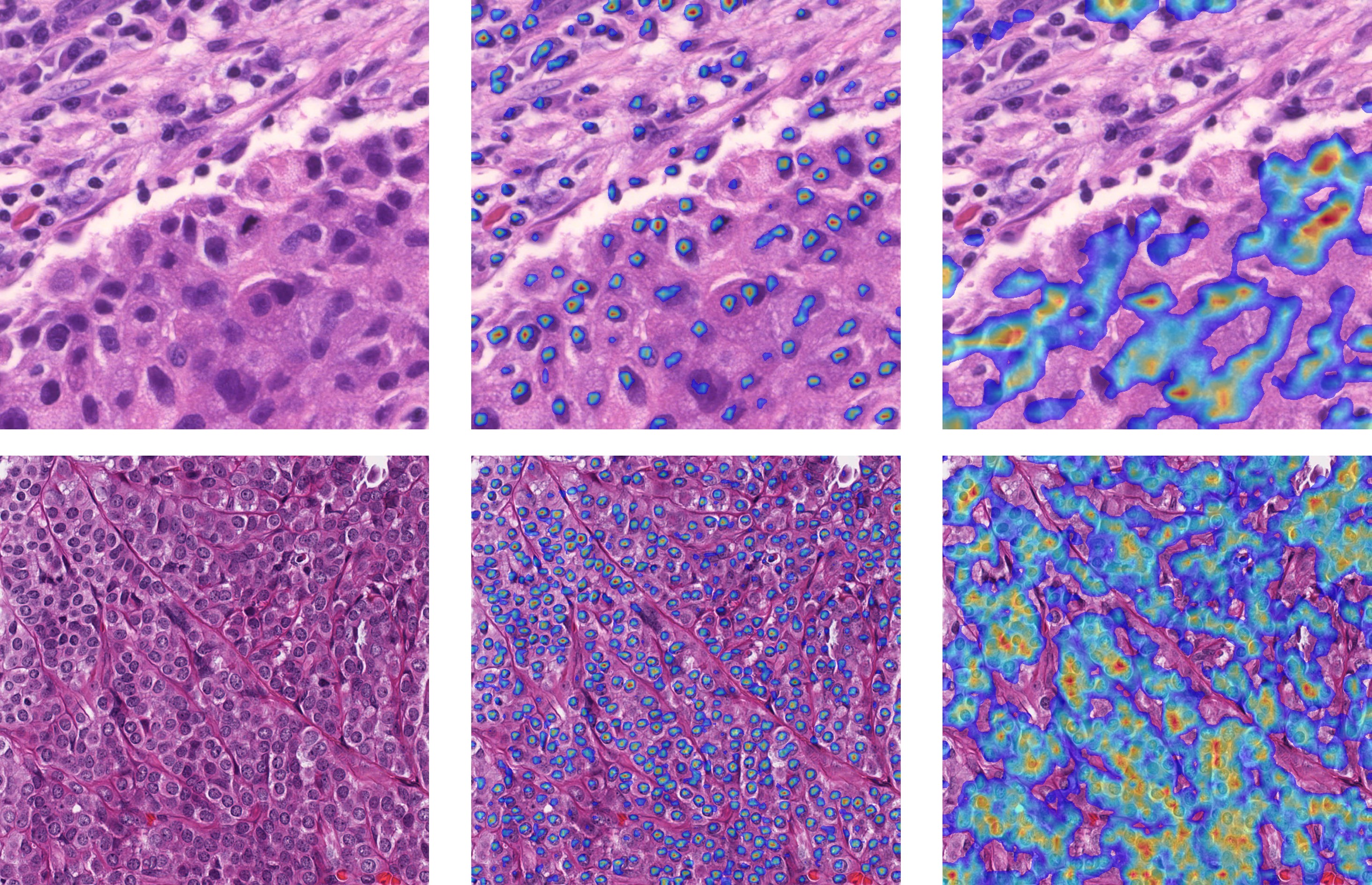}
	\caption{Input image (left) and corresponding Grad-CAM++ attention maps of context-free (middle) and context-aware (right) P2PNet models. It can be observed that incorporating high-level contextual features dilutes the model's attention of local nuclear morphological details.
	}\label{fig:heatmap}
\end{figure}

\subsection{Revitalizing Nuclear Morphological Perception}
Current context-aware nucleus detection methods primarily focus on incorporating richer tissue context to improve model performance \cite{shui2024dpa}. However, we discover that integrating high-level contextual features inevitably dilutes the model's attention of fine-grained nuclear morphological details, as depicted in Fig.~\ref{fig:heatmap}. This degraded perception could compromise the model's performance, as nuclear shape, size, chromatin pattern and texture are critical for accurate nucleus classification \cite{horst2024cellvit}.

Given that the segmentation task naturally models nuclear morphology, we utilize the auxiliary model developed in the previous section to alleviate this limitation. Specifically, we extract morphology-rich $m$ from the input feature maps of the model's last layer, which are specifically optimized to delineate nuclei regions and thereby contain abundant nuclear morphological information \cite{chen2023exploring}. Finally, we replace the input of $\phi^\prime$ from $e$ to $[e; m]$, leveraging complementary merits of nuclear contextual and morphological features for accurate classification.

During inference, we first feed $e$ to $\phi$ to identify foreground nuclei, then feed $[e;m]$ of each detected nucleus to $\phi^\prime$ for category prediction.

\begin{table*}[t!]
	\resizebox{0.99\linewidth}{!}{
		\begin{tabular}{c|cccc|ccc|cccc}
			\toprule[1.5pt]
			\multirow{2}{*}{Method} &
			\multicolumn{4}{c|}{BRCA} &
			\multicolumn{3}{c|}{OCELOT} &
			\multicolumn{4}{c}{PUMA} \\
			& $F^{Inf.}$ & $F^{Tum.}$ & $F^{Str.}$ & $F_{avg}$ & $F^{Tum.}$ & $F^{Back.}$ & $F_{avg}$ & $F^{Tum.}$ & $F^{Tils.}$ & $F^{Oth.}$ & $F_{avg}$ \\
			\midrule
			Hover-net & \score{62.31}{0.91} & \score{75.25}{0.24} & \score{49.58}{0.36} & \score{62.38}{0.22} & \score{69.27}{0.08} & \score{61.03}{0.46} & \score{65.15}{1.57} & \score{78.36}{0.24} & \score{75.72}{0.15} & \score{49.53}{0.16} & \score{67.87}{0.13} \\
			MCSpatNet & \score{64.45}{0.61} & \score{79.34}{0.28} & \score{55.31}{0.39} & \score{66.37}{0.17} & \score{70.69}{0.39} & \score{56.37}{1.23} & \score{63.53}{0.78} & \score{83.30}{0.16} & \score{77.74}{0.55} & \score{52.76}{1.31} & \score{71.27}{0.39} \\
			P2PNet & \score{64.25}{0.50} & \score{79.21}{0.15} & \score{55.20}{0.28} & \score{66.22}{0.10} & \score{72.55}{0.27} &  \score{61.64}{0.21} & \score{67.09}{0.23} & \score{83.35}{0.16} & \score{80.55}{0.13} & \score{56.93}{0.37} & \score{73.61}{0.17} \\
			Semi-P2PNet & \score{64.72}{0.73} & \score{81.11}{0.30} & \score{57.40}{0.33} & \score{67.74}{0.38} & \score{73.12}{0.36} & \score{62.03}{0.05} & \score{67.58}{0.20} & \score{83.69}{0.13} & \underline{\score{80.56}{0.34}} & \score{57.77}{0.27} & \score{74.01}{0.07} \\
			AC-Former & \score{59.77}{2.02} & \score{74.67}{0.33}  & \score{48.15}{1.59} & \score{60.87}{0.03} & \score{69.84}{0.13} & \score{58.74}{0.40} & \score{64.29}{0.20} & \score{79.51}{0.11} & \score{74.86}{0.48} & \score{54.15}{0.53} & \score{69.51}{0.20} \\
			SMILE & \underline{\score{71.82}{0.41}} & \score{80.27}{0.12} & \score{52.11}{0.07} & \score{68.06}{0.13} & \score{70.15}{0.16} & \score{61.08}{0.49} & \score{65.62}{0.31} & \score{79.89}{0.46} & \score{80.34}{0.18} & \score{53.10}{0.32} & \score{71.11}{0.27} \\
			PointNu-Net & \score{70.95}{0.51} & \score{80.04}{0.27} & \score{54.16}{0.16} & \score{68.38}{0.29} & \score{70.10}{0.29} & \score{59.01}{0.30} & \score{64.55}{0.28} & \score{81.03}{0.12} & \score{76.26}{0.16} & \score{53.53}{0.30} & \score{70.27}{0.18} \\
			CellViT & \score{68.12}{0.60} & \score{78.60}{0.47} & \score{50.27}{0.75} & \score{65.66}{0.49} & \score{70.25}{0.51} & \score{60.78}{0.64} & \score{65.52}{0.48} & \score{81.56}{0.24} & \score{78.85}{0.41} & \score{57.04}{0.68} & \score{72.49}{0.21} \\
			SENC & \score{56.89}{0.33} & \score{76.95}{0.12} & \score{50.31}{0.07} & \score{61.38}{0.17} & \score{73.70}{0.09} & \underline{\score{63.95}{0.23}} & \score{68.83}{0.16} & \score{79.87}{0.77} & \score{76.38}{0.60} & \score{53.14}{0.71} & \score{69.80}{0.69} \\
			CGT & \score{56.50}{0.11} & \score{76.61}{0.02} & \score{51.80}{0.05} & \score{61.63}{0.05} & \score{72.08}{0.06} & \score{62.80}{0.13} & \score{67.44}{0.09} & \score{79.89}{0.05} & \score{76.69}{0.02} & \score{54.51}{0.05} & \score{70.36}{0.03} \\
			TopoCellGen & \score{65.20}{0.40} & \underline{\score{81.70}{0.60}} & \underline{\score{58.20}{0.50}} & \underline{\score{68.40}{0.40}} & - & - & - & - & - & - & - \\
			MFoVCE-Net & \score{60.30}{1.04} & \score{78.79}{0.07} & \score{55.77}{1.77} & \score{64.95}{0.22} & \score{72.81}{0.59} & \score{63.17}{0.15} & \score{67.99}{0.37} & \score{79.80}{0.84} & \score{75.96}{1.13} & \score{55.16}{1.17} & \score{70.31}{0.84} \\
			MFoV-P2PNet & \score{63.52}{0.59} & \score{80.71}{0.24} & \score{55.84}{0.57} & \score{66.69}{0.37} & \underline{\score{74.70}{0.09}} & \score{63.48}{0.08} & \underline{\score{69.09}{0.06}} & \underline{\score{84.17}{0.29}} & \score{79.96}{0.54} & \underline{\score{59.70}{0.60}} & \underline{\score{74.61}{0.28}} \\
			\textbf{Ours} & \textbf{\score{72.68}{0.19}} & \textbf{\score{83.49}{0.10}} & \textbf{\score{59.87}{0.27}} & \textbf{\score{72.01}{0.13}} &  \textbf{\score{75.24}{0.28}} & \textbf{\score{66.43}{0.10}} & \textbf{\score{70.83}{0.15}} & \textbf{\score{86.45}{0.15}} & \textbf{\score{82.17}{0.29}} & \textbf{\score{63.45}{0.62}} & \textbf{\score{77.36}{0.30}} \\
			\bottomrule[1.5pt]
	\end{tabular}}
	\caption{Comparison of nucleus detection performance across three benchmarks. $F^{Inf.}$, $F^{Tum.}$, $F^{Str.}$, $F^{TILs.}$ and $F^{Oth.}$ denote the F1-socre for the inflammatory, tumor, stromal, TILs and other nuclei, respectively. $F_{avg}$ represents the average F1-score. The best and second-best performance are highlighted in \textbf{bold} and \underline{underlined}, respectively.
	}\label{tab:det}
\end{table*}
\begin{table*}[t!]
	\resizebox{0.99\linewidth}{!}{
		\begin{tabular}{c|cccc|ccc|cccc}
			\toprule[1.5pt]
			\multirow{2}{*}{Method} &
			\multicolumn{4}{c|}{BRCA} &
			\multicolumn{3}{c|}{OCELOT} &
			\multicolumn{4}{c}{PUMA} \\
			& $PQ^{Inf.}$ & $PQ^{Tum.}$ & $PQ^{Str.}$ & $PQ_{avg}$ & $PQ^{Tum.}$ & $PQ^{Back.}$ & $PQ_{avg}$ & $PQ^{Tum.}$ & $PQ^{Tils.}$ & $PQ^{Oth.}$ & $PQ_{avg}$ \\
			\midrule
			Hover-net & \score{46.97}{0.69} & \score{64.58}{0.16} & \score{45.52}{0.27} & \score{52.36}{0.19} & \score{61.15}{0.40} & \score{43.44}{0.63} & \score{52.29}{0.29} & \score{64.17}{0.18} & \score{56.36}{0.06} & \score{35.17}{0.19} & \score{51.90}{0.09} \\
			MCSpatNet & \score{51.24}{0.93} & \score{69.66}{0.18} & \score{44.65}{0.37} & \score{55.18}{0.38} & \score{65.89}{0.18} & \score{38.88}{1.30} & \score{52.39}{0.65} & \score{66.74}{0.21} & \score{55.30}{0.45} & \score{31.96}{1.40} & \score{51.33}{0.54} \\
			P2PNet & \score{49.92}{1.46} & \score{69.15}{0.18} & \score{46.74}{0.14} & \score{55.27}{0.48} & \score{66.21}{0.28} & \score{44.44}{0.36} & \score{55.33}{0.28} & \score{67.16}{0.17} & \score{58.11}{0.25} & \textbf{\score{39.95}{0.29}} & \score{54.74}{0.17} \\
			Semi-P2PNet & \score{48.93}{0.44} & \underline{\score{71.17}{0.24}} & \underline{\score{47.69}{0.16}} & \score{55.93}{0.19} & \score{66.70}{0.38} & \score{44.76}{0.95} & \score{55.73}{0.67} & \score{67.13}{0.29} & \underline{\score{58.40}{0.12}} & \score{39.10}{0.36} & \underline{\score{54.88}{0.18}} \\
			AC-Former & \score{42.54}{0.17} & \score{63.44}{1.52} & \score{40.87}{0.32} & \score{48.95}{0.34} & \score{64.97}{0.41} & \score{42.87}{0.53} & \score{53.92}{0.16} & \score{64.78}{0.14} & \score{55.11}{0.44} & \score{34.58}{0.26} & \score{51.49}{0.21} \\
			SMILE & \score{45.71}{0.94} & \score{70.95}{0.24} & \score{44.66}{0.18} & \score{53.77}{0.33} & \score{63.07}{0.22} & \score{45.56}{0.64} & \score{54.32}{0.29} & \score{65.44}{0.43} & \score{57.27}{0.17} & \score{37.44}{0.46} & \score{53.38}{0.14} \\
			PointNu-Net & \score{50.34}{0.25} & \score{71.16}{0.33} & \score{46.78}{0.13} & \underline{\score{56.09}{0.14}} & \score{64.08}{0.28} & \score{43.95}{0.27} & \score{54.01}{0.23} & \score{66.76}{0.23} & \score{54.76}{0.12} & \score{36.64}{0.54} & \score{52.72}{0.28} \\
			CellViT & \score{45.52}{1.07} & \score{70.59}{0.19} & \score{43.51}{0.36} & \score{53.21}{0.43} & \score{64.46}{0.69} & \underline{\score{46.13}{0.91}} & \score{55.29}{0.57} & \score{64.68}{0.33} & \score{54.24}{0.89} & \score{36.24}{0.54} & \score{51.72}{0.46} \\
			SENC & \score{36.98}{0.17} & \score{67.86}{0.14} & \score{41.44}{0.11} & \score{48.76}{0.11} & \score{66.77}{0.04} & \score{44.67}{0.24} & \score{55.72}{0.13} & \score{64.88}{0.66} & \score{51.75}{0.19} & \score{33.05}{0.22} & \score{49.89}{0.34} \\
			CGT & \score{38.18}{0.05} & \score{67.98}{0.02} & \score{42.26}{0.04} & \score{49.47}{0.02} & \score{63.08}{0.11} & \score{41.87}{0.09} & \score{52.47}{0.02} & \score{65.08}{0.04} & \score{53.83}{0.03} & \score{34.46}{0.10} & \score{51.12}{0.03} \\
			MFoVCE-Net & \underline{\score{51.46}{1.49}} & \score{69.09}{1.13} & \score{45.86}{1.26} & \score{55.47}{0.54} & \score{66.74}{0.07} & \score{44.42}{0.14} & \score{55.58}{0.09} & \score{65.26}{0.38} & \score{55.10}{1.10} & \score{34.94}{0.69} & \score{51.76}{0.26} \\
			MFoV-P2PNet & \score{50.01}{1.79} & \score{70.85}{0.23} & \score{46.63}{0.56} & \score{55.83}{0.68} & \underline{\score{67.18}{0.29}} & \score{45.52}{0.16} & \underline{\score{56.35}{0.15}} & \underline{\score{68.42}{0.24}} & \score{57.83}{0.35} & \score{38.23}{0.54} & \score{54.83}{0.28} \\
			\textbf{Ours} & \textbf{\score{54.82}{0.54}} & \textbf{\score{73.30}{0.25}} & \textbf{\score{49.24}{0.24}} & \textbf{\score{59.12}{0.18}} & \textbf{\score{67.62}{0.29}} & \textbf{\score{48.85}{0.36}} & \textbf{\score{58.24}{0.29}} & \textbf{\score{69.53}{0.13}} & \textbf{\score{58.53}{0.28}} & \underline{\score{39.92}{0.22}} & \textbf{{\score{55.99}{0.13}}} \\
			\bottomrule[1.5pt]
	\end{tabular}}
	\caption{Comparison of nucleus instance segmentation performance across three benchmarks. To enable nucleus detection models to produce instance masks, we train the segmentor component of PromptNucSeg \cite{shui2024unleashing} on the training sets of these datasets and employ different detection models as the prompter component within the PromptNucSeg framework.}\label{tab:seg}
\end{table*}

\subsection{Experiments}
\subsection{Experimental Setup}
\textbf{Datasets.}
To the best of our knowledge, there are currently no publicly available datasets with exhaustive WSI-level nucleus annotations. Consequently, we conduct experiments on three patch-level benchmarks that provide context images for each annotated patch.
\begin{itemize}
	\item \textbf{BRCA} \cite{abousamra2021multi} is a breast cancer dataset and consists of 120 patches at 20$\times$ magnification belonging to 113 patients, collected from TCGA \cite{weinstein2013cancer}. The training, validation, and testing sets contain 80, 10, and 30 patches, respectively. The nuclei in this dataset are categorized into three types: tumor, inflammatory, and stromal.
	\item \textbf{OCELOT} \cite{ryu2023ocelot} comprises 664 patches at 40$\times$ magnification extracted from 303 WSIs. The dataset contains a total of 113,026 nuclei, annotated to differentiate between tumor and non-tumor nuclei. The training, validation, and test sets contain 400, 137, and 126 patches, respectively.
	\item \textbf{PUMA} \cite{shahamiri2025cracking} comprises 206 patches extracted from melanoma tissue scanned at 40× magnification. It provides annotations for three nuclei types: tumor, tumor infiltrating lymphocytes  (TILs), and other nuclei. We randomly divide this dataset into training, validation, and test subsets at a ratio of 6:2:2.
\end{itemize}

Additionally, we manually annotate instance masks for nuclei in the BRCA and OCELOT datasets. These annotations serve two purposes: enabling the training of baseline methods like CellViT \cite{horst2024cellvit} that require mask supervision and facilitating the evaluation of different methods on context-aware nucleus instance segmentation.
%	The PUMA dataset requires no additional processing as it already provides instance mask annotations.

\noindent\textbf{Evaluation metrics.} For nucleus detection, following previous studies \cite{abousamra2021multi,ryu2023ocelot}, we employ F1-score as the evaluation metric. If a detected nucleus is within a valid distance ($\sigma$) from an annotated nucleus and the nuclear class matches, it is counted as a true positive (TP), otherwise a false positive (FP). Then, the F1-score for the $c$-th class is calculated as: $\rm{F}_c = \frac{2 \rm{TP_c}}{2\rm{TP_c} + \rm{FP_c} + \rm{FN_c}}$, and the average F1-score is $\bar{F} = \frac{1}{C} \sum_{c=1}^{C} \rm{F}_c$. In accordance with the official settings, $\sigma$ is set to 6, 15 and 15 pixels for the BRCA, OCELOT and PUMA benchmarks, respectively. For nucleus instance segmentation, we adopt Panoptic Quality (PQ) \cite{kirillov2019panoptic} as the evaluation metric. Following \cite{ryu2023ocelot}, we repeat all experiments for 5 times with different random seeds and report the mean and 95\% confidence interval of the performance metrics.

\begin{figure*}[t!]
	\centering
	\includegraphics[width=0.99\linewidth]{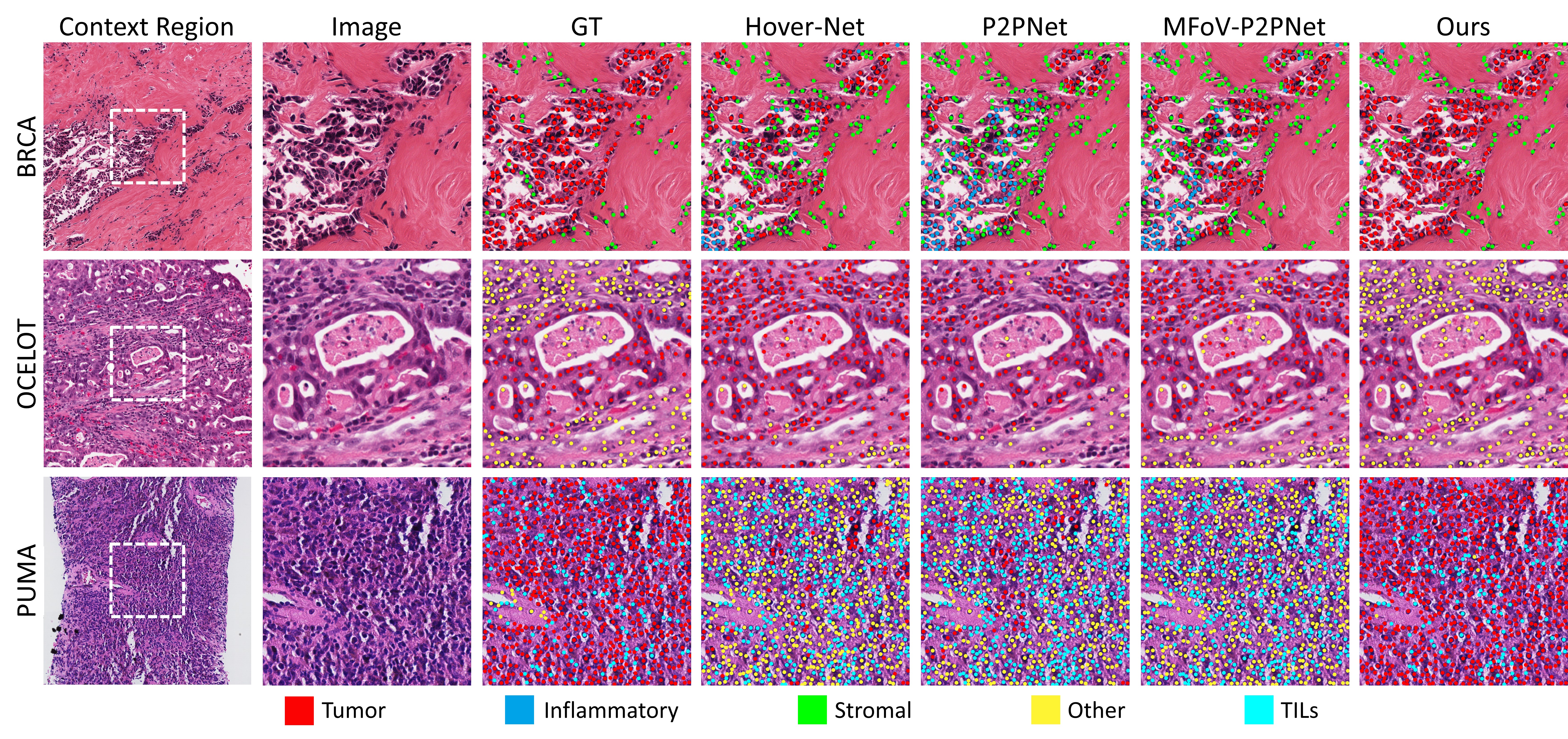}
	\caption{Qualitative comparison on three benchmarks. The white dashed boxes indicate the positions of ROI detection patches within the context images.}\label{fig:vis}
\end{figure*}

\noindent\textbf{Implementation details.} We use ResNet-50 \cite{he2016deep} pre-trained on ImageNet-1K as the image encoder. The detector is trained using the AdamW \cite{loshchilov2017decoupled} optimizer with learning rate of 1e-4 for 200 epochs. We use a batch size of 2, with all data distributed across 4 V100 GPUs. The auxiliary cross-labeling model comprises 12 convolutional blocks \cite{liu2022convnet}, FPN \cite{lin2017feature} and two PixelShuffle layers. It accounts for 9\% of the total network parameters and is trained for 20 epochs. The head $\phi^\prime$ consists of two linear layers and is trained for 100 epochs with cross-entropy loss. We set $k=3$ and pooling size $o=6$ in our experiments. Unless otherwise specified, the size of context area $\delta$ is set to 1.

\begin{table}[t!]
	\centering
	\resizebox{0.99\linewidth}{!}{
		\begin{tabular}{c|cccc}
			\toprule[1.5pt]
			Method & Params (M) & Time (s) & FLOPs (G) & $F_{avg}$ \\
			\hline
			CellViT & 142.85 & 3027.04 & 3566.71 & 65.66 \\
			P2PNet & \textbf{27.26} & \textbf{148.86} & \textbf{100.08} & 66.22 \\
			MFoV-P2PNet & 53.22 & 486.20 & 212.75 & 66.69 \\
			Ours w/ ME & 48.08 & 205.81 & 186.74 & \textbf{72.01} \\
			Ours w/o ME & \underline{44.08} & \underline{156.07} & \underline{115.28} & \underline{71.23} \\
			\bottomrule[1.5pt]
	\end{tabular}}
	\caption{Comparison on model size, computational cost, inference efficiency and performance of different methods. The inference time is measured using ten TCGA-BRCA WSIs with an average of 5k sliding window patches of size 1024$\times$1024. ME represents retaining the auxiliary model to supplement nuclear morphology-rich embedding.}
	\label{tab:speed}
\end{table}

\subsection{Comparison with the State-of-the-art Methods}
We compare the proposed method with state-of-the-art (SOTA) nucleus instance segmentation methods including Hover-Net \cite{graham2019hover}, SMILE \cite{pan2023smile}, PointNu-Net \cite{yao2024pointNu} and CellViT \cite{horst2024cellvit}, as well as nucleus detection methods, including MCSpatNet \cite{abousamra2021multi}, P2PNet \cite{song2021rethinking}, Semi-P2PNet \cite{shui2023semi}, CGT \cite{lou2024cell}, SENC \cite{lou2024structure}, TopoCellGen \cite{xu2025topocellgen}, MFoVCE-Net \cite{bai2020multi}, and MFoV-P2PNet \cite{shui2024dpa}. Among these methods, MFoVCE-Net and MFoV-P2PNet are context-aware approaches, while the others are context-free. Semi-P2PNet is a semi-supervised learning approach that also utilizes unlabeled nucleus samples in surrounding patches.

Tab.~\ref{tab:det} and Tab.~\ref{tab:seg} exhibit the performance comparison results on the detection and segmentation tasks, respectively. For nucleus detection, the proposed method outperforms the SOTA counterparts by 3.61, 1.74 and 2.75 points in $F_{avg}$ on the BRCA, OCELOT and PUMA benchmarks. Compared to the baseline P2PNet model, our method shows substantial improvements of 5.79, 3.74 and 3.75 points on average F1 score. In task of nucleus instance segmentation, our method achieves 3.03, 1.89 and 1.11 points improvement over the SOTA methods in $PQ_{avg}$ across three datasets.

Fig.~\ref{fig:vis} presents qualitative comparison results on three benchmarks. Taking the image from the BRCA dataset as an example, the leftmost image shows the context region, where a cluster of densely packed, hyperchromatic nuclei is clearly visible in the lower left corner and displays a tumor invasion area. Under sliding window inference, without access to this broader contextual information, context-free nucleus detection models incorrectly classify many of the tumor nuclei as inflammatory or stromal types. Among context-aware approaches, our method exhibits better nucleus detection accuracy than MFoV-P2PNet. This can be attributed that compared to MFoV-P2PNet that uses LFoV patches to extract contextual information, our approach leverages surrounding window patches with higher magnification that provide more detailed contextual features.

Tab.~\ref{tab:speed} presents a comprehensive comparison of model size, computational cost, inference efficiency and performance across different methods. All metrics are measured on a system with a single RTX 3090 GPU and dual AMD EPYC 7542 processors (2.90GHz, 64 cores, 128 threads). Compared to the baseline context-free nucleus detector (\ie, P2PNet), our method introduces minimal additional inference time. Notably, the proposed model runs 2.36$\times$ faster than previous context-aware detector (\ie, MFoV-P2PNet) as we eliminate the time-consuming step of additionally preparing LFoV patches.

\subsection{Ablation study}
We evaluate the effect of our proposed modules and hyper-parameters on the BRCA dataset. Due to page limitation, we report only mean results across five runs.

\noindent\textbf{Effect of our proposed modules}.
Tab.~\ref{tab:abl} presents the ablation results of context-aware learning (CA), cross-labeling (CL) and the incorporation of morphology-rich embedding (ME). It can be observed that all proposed modules contribute to improving the model performance.

\noindent\textbf{Effect of $\delta$.} Tab.~\ref{tab:delta} shows the impact of context size $\delta$ on model performance. $\delta=0$ denotes the baseline P2PNet model. The results show dramatic performance gains when $\delta$ increases from 0 to 1, whereas further increments yield marginal improvements. This is also observed with previous context-aware method, MFoV-P2PNet. We hypothesize that the nearest neighboring patches provide the most relevant contextual information, while incorporating broader context regions inevitably introduces background noise that impedes the model from identifying informative contextual features.
%	To mitigate this challenge, we plan to introduce a distance-based weighting scheme in future work.

\begin{table}[t!]
	\centering
	\begin{minipage}{0.23\textwidth}
		\centering
		\begin{tabular}{ccc|c}
			\toprule[1.5pt]
			CA & CL & ME & $F_{avg}$ \\
			\hline
			& & & 66.22 \\
			\cmark & & & 70.79 \\
			\cmark & & \cmark & 70.95 \\
			\cmark & \cmark & & 71.23 \\
			\cmark & \cmark & \cmark & 72.01 \\
			\bottomrule[1.5pt]
		\end{tabular}
		\caption{Effect of our proposed modules.}\label{tab:abl}
	\end{minipage}
	\hfill
	\begin{minipage}{0.23\textwidth}
		\centering
		\begin{tabular}{lcc}
			\toprule[1.5pt]
			$\delta$ & MFoV-P2PNet & Ours \\
			\hline
			0 & 66.22 & 66.22 \\
			1 & 66.69 & 72.01 \\
			2 & 66.83 & 72.07 \\
			3 & 66.95 & 72.28 \\
			4 & 66.98 & 72.51 \\
			\bottomrule[1.5pt]
		\end{tabular}
		\captionof{table}{Effect of context size $\delta$.}
		\label{tab:delta}
	\end{minipage}
\end{table}

\begin{figure}[t!]
	\centering
	\begin{minipage}{0.22\textwidth}
		\centering
		\begin{tabular}{lcc}
			\toprule[1.5pt]
			Method & $F_{avg}$ & $\Delta$ \\
			\midrule
			baseline & 70.95 & - \\
			SL & 71.10 & +0.15 \\
			CL & 72.01 & +1.06 \\
			\bottomrule[1.5pt]
		\end{tabular}
		\captionof{table}{Effect of pseudo-labeling strategies.}
		\label{tab:abl_labeling}
	\end{minipage}
	\hfill
	\begin{minipage}{0.22\textwidth}
		\centering
		\begin{tabular}{lc}
			\toprule[1.5pt]
			Method & $F_{avg}$ \\
			\hline
			Add & 67.50 \\
			Concat & 68.85 \\
			CrossAttn & 72.01 \\
			\bottomrule[1.5pt]
		\end{tabular}
		\captionof{table}{Effect of context integration strategies.}
		\label{tab:integration}
	\end{minipage}
\end{figure}

\begin{figure}[t!]
	\centering
	\includegraphics[width=\linewidth]{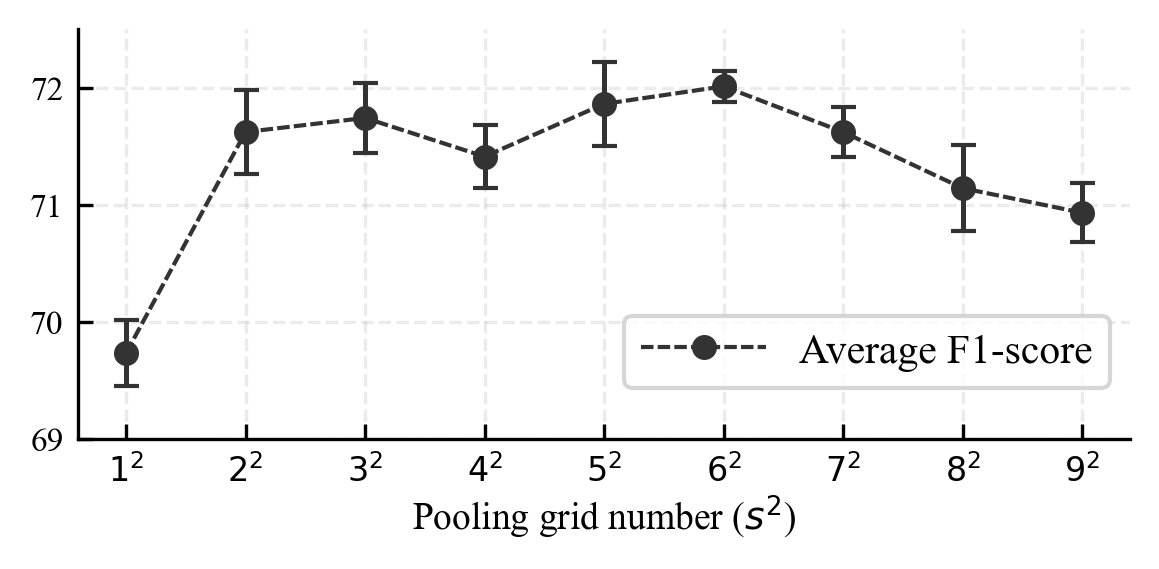}
	\caption{Effect of $s$. The error bars denote 95\% confidence intervals across 5 independent runs.}\label{fig:pooling_size}
\end{figure}

\noindent\textbf{Comparison of pseudo-labeling strategies}.
Tab.~\ref{tab:abl_labeling} shows the model performance with different pseudo-labeling strategies. The baseline represents the model with CA and ME modules. It can be observed that self-labeling (SL), which uses the detector itself to generate pseudo class labels for nuclei in surrounding patches, yields marginal performance gains. The proposed cross-labeling (CL) strategy that utilizes another model to produce pseudo labels shows notable improvement. Although our evaluations show that the auxiliary model does not exhibit better classification accuracy than the base detector (70.74 vs 70.79), it effectively mitigates the error accumulation problem inherent in the SL strategy.

\noindent\textbf{Comparison of context integration strategies}. Tab.~\ref{tab:integration} shows the effect of different context integration methods. The cross-attention (CrossAttn) operation achieves significantly better results than both addition (Add) and concatenation approaches (Concat).

\noindent\textbf{Effect of $s$}. Fig.~\ref{fig:pooling_size} exhibits the impact of pooling grid number on model performance, with optimal results achieved at $s=6$. Intuitively, increasing the grid number preserves more comprehensive contextual details and leads to better results. However, we observe performance degradation when $s$ exceeds 6. This can be attributed to the inherent spatial redundancy in high-level contextual feature maps \cite{li2023scconv,yang2025visionzip}, where the feature points have already undergone sufficient information exchange in the earlier stages. This redundancy may impede the model in capturing informative contextual features and thus resulting in sub-optimal performance. 

\section{Conclusion}
In this paper, we propose an effective and efficient context-aware nucleus detection approach. Instead of using LFoV patches in previous methods, we aggregate contextual information from historically visited sliding windows during whole-slide inference, which significantly improves the detection efficiency and accuracy. Additionally, we propose a cross-labeling strategy to effectively leverage unlabeled nuclei in surrounding patches to enhance the model's context adaptability, and incorporate an auxiliary model to compensate for the loss of nuclear morphological details during contextual features integration. Extensive experiments on three benchmarks demonstrate the superiority of our method.

\noindent\textbf{Limitations and future work}.
Compared to previous context-aware methods, our approach suffers from increased training time, as each iteration requires encoding a substantially larger number of patches. To address this limitation, we plan to explore feature caching strategies to accelerate model training in future work.

\section{Acknowledgements}

This study was partially supported by "Pioneer" and "Leading Goose" R\&D Program of Zhejiang (Grant 2025SDXHDX0003), the National Natural Science Foundation of China (Grant No.62506306), and foundation of Muyuan Laboratory (Program ID: 14106022401,14106022402). Furthermore, essential technical support was provided by the D-PathAI platform, including its hardware and software, which was developed by Hangzhou Dipath Technology Co., Ltd.

\bibliography{aaai2026}

\end{document}